\documentclass[reprint,amsmath,amssymb,aps,prl,superscriptaddress]{revtex4-1}
\usepackage{hyperref}
\usepackage[dvipsnames]{xcolor}%To use extra colors
\usepackage{graphicx}% Include figure files
\usepackage{dcolumn}% Align table columns on decimal point
\usepackage{bm}% bold math
\usepackage{color}
\usepackage{bbold}
\usepackage{soul}
\usepackage{epsfig,epic}
\usepackage{epstopdf}
\usepackage{nicefrac}
\usepackage{verbatim}
\usepackage{amsthm}
\newcommand{\newc}{\newcommand}
\newc{\beq}{\begin{equation}}
\newc{\eeq}{\end{equation}}
\newc{\kt}{\rangle}
\newc{\br}{\langle}
\newc{\beqa}{\begin{eqnarray}}
\newc{\eeqa}{\end{eqnarray}}
\newc{\tr}{\mbox{tr}}
\newc{\ovl}{\overline}
\newc{\longra}{\longrightarrow}

\newtheorem*{theorem}{Theorem}

\newtheorem*{corollary}{Corollary}

\hypersetup{
    colorlinks=true,
    linkcolor=blue,
    filecolor=magenta,      
    urlcolor=blue,
    citecolor= blue
}
\makeatletter
\let\Hy@backout\@gobble
\makeatother

\begin{document}

%\title{Enhancing nonlocality with Intermediate local operations}
%\title{Exponential approach to typicality with random local unitary operators}
\title{Can local dynamics enhance entangling power?}

\author{ Bhargavi Jonnadula}
\affiliation{Department of Physics, Indian Institute of Technology Madras, Chennai, India~600036}
\author{Prabha Mandayam}
\affiliation{Department of Physics, Indian Institute of Technology Madras, Chennai, India~600036}
\author{Karol \.Z{}yczkowski}
%\email{karol@tatry.if.uj.edu.pl}
\affiliation{Smoluchowski Institute of Physics, Jagiellonian University, Cracow, Poland}
\affiliation{Center for Theoretical Physics, Polish Academy of Sciences, Warsaw, Poland}
\author{Arul Lakshminarayan}
\affiliation{Department of Physics, Indian Institute of Technology Madras, Chennai, India~600036}

%\affiliation{}
%\affiliation{}
%\author{}
%\affiliation{}

\date{Oct. 22, 2016}

%%%Short abstract
%\begin{abstract}
%Yes.
%\end{abstract}

%%%%Long Abstract
\begin{abstract}
It is demonstrated here that local dynamics have the ability to strongly modify %, sometimes increase, 
the entangling power of unitary quantum gates acting on a composite system.
The scenario is common to numerous physical systems, in which the
time evolution involves local operators and nonlocal interactions.
%%"one-body" removed.
To distinguish between distinct classes of gates with zero entangling power
we introduce a complementary quantity called gate-typicality
and study its properties.
Analyzing multiple applications of any entangling operator interlaced with random local gates, we prove
that both investigated quantities approach their asymptotic values 
in a simple exponential form. This rapid convergence to equilibrium,
valid for subsystems of arbitrary size,
is illustrated by studying multiple actions of 
diagonal unitary gates and controlled unitary gates. 
\end{abstract}

%\pacs{PACS here}

\maketitle

\noindent {\it Introduction}: 
The uniquely nonclassical phenomenon of entanglement is a well-known resource for quantum information \cite{NielsenChuang}.  It is increasingly used to characterize complex states, from many-body ground states \cite{Amico08},
to infinite temperature quantum phase transitions such as the ergodic 
to localized phase in strongly interacting many-body  systems \cite{Kjaell14}. 
Simple coupled quantum chaotic models have been studied \cite{Lakshminarayan01, Trail08} and  experimentally realized \cite{Chaudhury2009,Neill16} to demonstrate the large entanglement growth wherein the subsystems are nearly maximally mixed.
In general the dynamics of entanglement in a non-equilibrium context is responsible for thermalization \cite{Linden2009}.  More recently in a different setting, black-holes are conjectured to scramble quantum information in a time that is logarithmic in the entropy via entanglement \cite{Lashkari2013}.
%An intriguing connection of entanglement growth and spread in noisy spatially local systems with unitary dynamics and the KPZ random surface growth has recently been proposed. }

While much work has centered on properties of states, quantum operators have also been studied as a physical resource for creating entanglement  \cite{Zanardi2000,Collins2001, Vidal2002,Hammerer2002,Nielsen2003,Emerson2003}.  
Studying directly the operators, such as propagators in time, frees us from the arbitrariness of initial states or the choice of eigenvectors. The entangling power \cite{Zanardi2000,Zanardi2001} while referring to an inherent property of operators on a bipartite composite system is also related to how much state entanglement can be created, on the average, using one application of the unitary on product states. Investigations on quantum transport in light harvesting
complexes \cite{Caruso2010}, quantum chaos \cite{Dobrzanski2004} 
and thermalization \cite{Scarani2002}
have  made direct use of the entangling power of bipartite unitary gates.

This Letter concerns the evolution of entangling power when multiple nonlocal operators are used successively while being interlaced with local operators, a typical situation in time evolution.
Local unitary invariance, in a bipartite setting, implies that $(U^A \otimes U^B) \, U \,({U^A}^\prime \otimes {U^B}^\prime)$ has the same  entangling power as  $U$. However, if nonlocal operations are interlaced by local dynamics, the consequences are nontrivial, as the entangling power of $U$ and $\sqrt{U} \,( U^A\otimes U^B)\, \sqrt{U}$ are not the same.

 Specifically, we are interested in the case when the nonlocal operators 
%,including controlled gates, 
are fixed and structured, while  complexity is introduced in local gates taken randomly
from a given ensemble. The resultant operators are shown to rapidly acquire properties of random operators on the composite space, in particular, large entangling powers are obtained however small the entangling power of the individual nonlocal operators may be.
The resulting entangling power can be, counterintuitively, 
 much larger than what can be achieved in the absence of the local operators.
 Besides the importance of such scenarios in the context of dynamical evolution,
 they indicate how generic bipartite gates can be prepared using 
 local random operators  \cite{Emerson2003}.
% albeit in the simplest bipartite setting. 
It is known that such composite random operators allow for protocols such as approximate quantum encryption and data hiding that are more efficient than their deterministic counterparts \cite{Hayden2004}.
% Random quantum circuits are useful models of both approach to typical random unitary operators and for the speed of scrambling.}

%It is also known that any unitary gate entangling two qubits supplemented 
%by local gates is universal for quantum computation \cite{Brylinski2002,Bremner2002}.
%Thus characterizing the strength of quantum 
%operations acting on two subsystems 
%becomes indispensable while
%%forms a crucial step towards 
%allocating resources and measuring computational complexity.

%Furthermore, for a study of unitary time evolution of composite 
%quantum systems it is important to describe interaction
%between individual subsystems. 
%On the other hand, when the operator is the time propagator, this %could also be a means to study the evolution of complex quantum
% systems free from the specificity of states. Thus for example 

%The present work deals exclusively with unitary operators.

%Apart from entangling power, other measures of interaction strengths of unitary operators
%were introduced and characterized in \cite{Nielsen2003}. 
 We investigate the problem in detail, applying the entangling power $e_p(U)$ %, 
%% line below can be commented out %%%%
%-- a widely used measure of the interaction strength --
%%%%%%%%
and introducing a complementary quantity $g_t(U)$ that, unlike $e_p(U)$, differentiates between 
local gates and the swap gate. Access to long-time or multiple uses of nonlocal operators is 
possible by analytically averaging over an ensemble of random local gates.
Explicit results obtained in this way are shown to provide an excellent approximation 
for the time dependence of both the entangling power and $g_t$ of multiple usage of a given bipartite unitary gate.

\noindent {\it Entangling power and gate-typicality:} Most measures of operator interaction strengths \cite{Nielsen2003} are based on the Schmidt decomposition of the unitary evolution operator $U$ acting 
on a bipartite space ${\cal H}^N\otimes {\cal H}^N$.
The operator Schmidt  decomposition and the ``operator entanglement'' $E(U)$ 
read \cite{Nielsen2003}
\beq 
\label{eqn:SchmidtEU}
U=\sum_{i=1}^{N^2}\sqrt{\lambda_i} \, A_i \otimes B_i,\ E(U)=1-\dfrac{1}{N^4}\sum_{i=1}^{N^2}\lambda_i^2.
\eeq
Here $A_i$ and $B_i$ are orthonormal operators, {\it i.e.} $\tr(A_i A_j^{\dagger})=\tr(B_i B_j^{\dagger})=\delta_{ij}$ and $\lambda_i\ge0$. Unitarity implies that $\sum_{i=1}^{N^2}\lambda_i = N^2$ and therefore the 
set  $\{\lambda_i/N^2,1\le i \le N^2  \}$ forms a discrete probability measure,
and $E(U)$ is  the operator linear entropy. 
%The Schmidt strength of an operator was defined via the von Neumann entropy in \cite{Nielsen2003}, however for the purposes of analytical calculations the linear entropy is preferred in the present work.

The operator entanglement of a unitary gate $E(U)$ is linked
to its average ability to create entanglement. 
The {\sl entangling power} of an operator $U$ is defined as  $e_p(U)=\overline{ E_L(U|\psi_{A}\kt|\psi_B\kt)}^{\psi_A,\psi_B}$ -- see
\cite{Zanardi2000,Zanardi2001,Wang2002}. 
Here  $E_L(|\psi_{AB}\kt)$ is the usual linear entropy of the state $|\psi_{AB}\kt$,
defined as $E_L=1-\tr \rho_A^2$, where $\rho_A$ is the reduced density matrix of $A$,
and the average is taken over all the product states $|\psi_A\kt |\psi_B\kt$
distributed according to the unitarily invariant measure.
Entangling power of a gate $U$ of size $N^2$ is bounded as:
 $0 \le e_p(U) \le (N-1)/(N+1)$.
Interestingly, both quantities are directly related
as shown by Zanardi in \cite{Zanardi2001}, 
$e_p(U)=N^2 [E(U)+E(US)-E(S)]/(N+1)^2$.
% with $0 \le e_p(U) \le (N-1)/(N+1)$.
Here $S$ is the {\sc swap} gate defined by $S|\psi_A\kt |\psi_B\kt =|\psi_B\kt |\psi_A\kt$,
so that $E(S)=(N^2-1)/N^2$.

Evaluation of $E(U)$ and $E(US)$, as well as their interpretation, is facilitated by considering them as entanglement measures of four-party pure states -- see Supplementary Material.
Towards this end, consider two density matrices
\beq
\label{eqn:rhoRT}
%\begin{split}
\rho_{R}(U)= \frac{1}{N^2} \,U_RU_R^{\dagger}, \ \ \ \  %\text{and}\, 
 \rho_{T}(U)=\frac{1}{N^2}S \,U_{T}U_{T}^{\dagger}S,
%\end{split}
\eeq
where  $U_R$ is the reshuffling of $U$, while $U_{T}$ is its partial transpose with respect to $A$. These are defined as the following, generally non-unitary, permutations of the original bi-partite
unitary matrix $U$: $\br ij |U_R|\alpha \beta \kt = \br i \alpha|U|j \beta\kt$, and $\br j \alpha |U_{T}| i \beta \kt = \br i \alpha|U|j \beta\kt$.

The operator Schmidt decomposition of $U$ is 
determined by the spectra of $U_R \, U_R^{\dagger}$ \cite{Zyczkowski2004}, 
%to be precise 
as the eigenvalues of $\rho_R(U)$ are equal to the rescaled 
coefficients $\lambda_i/N^2$ from  Eq.~\eqref{eqn:SchmidtEU}.
As $S(SU)_R=U_{T}$, it is easy to relate the eigenvalues of $\rho_{T}(U)$ to the Schmidt decomposition of $SU$. It follows that 
%\beq
%\label{eqn:EUESU}
$E(U)=1-\tr(\rho_{R}^2(U))$, and $E(SU)=1-\tr(\rho_{T}^2(U)).$
%\eeq
The operator linear entropy $E(U)$ is thus equal to the linear entropy of the state $\rho_{R}(U)$. 
Related observations previously appeared in \cite{Wang2003,Ma2007}. 
\medskip 

 While $E(U)$ and $E(US)$ are two independent polynomial invariants of $U$ \cite{Grassl1998}, 
the entangling power is a symmetric function of these quantities
and hence does not differentiate local dynamics from the {\sc swap} gate $S$. 
Therefore, it is useful to introduce a quantity complementary to the entangling power $e_p(U)$, 
defined by the antisymmetric combination 
\beq
\label{eqn:enlU}
g_{t}(U)=\dfrac{N^2}{N^2-1}[ E(U)-E(US) +E(S)].
\eeq

%Then $0 \le g_{t}(U) \le 2$ and it is shown below that 
%\beq
%\label{eqn:etaexp}
%\br g_{t}(U^{(n)}) \kt_W= 1-[1-g_{t}(U)]^n,
%\eeq
%is the local operators averaged nonlocality on $n$ applications of $U$.
% %For example if $U$ is {\sc swap} then $\br g_{t}(U^{(n)}) \kt=1-(-1)^n$, and is zero if there are an even number of {\sc swap} gates and is otherwise the maximum value of $2$. 
% As long as $g_{t}(U) \ne 0,\, \mbox{or}\, 2$, for $n \rightarrow \infty$, $\br g_{t}(U^{(n)}) \kt \rightarrow  1$ which is the typical value. 

%That $g_t(U)$ is unlike $E(U)$ itself, is reflected in the fact that 
Observe that the maximum value, $g_t=2$, is achieved for the {\sc swap} gate,
and locally equivalent gates, $(U^A \otimes U^B)\, S\, (U^{A'} \otimes U^{B'})$. 
% $U=S$, that is it is a {\sc swap} gate (up to multiplications by local operators).
This is consistent with the fact that, although $e_p(S)=0$, in terms of the operator entanglement $E(U)$, the  {\sc swap} gate is a maximally nonlocal operator as all its Schmidt coefficients $\lambda_i$ are equal. Operationally as well,   when implementing gates using teleportation and classical communication, the {\sc swap} gate consumes maximum resources \cite{Collins2001,Eisert2000}.
%%uncomment: \textcolor{blue}{Modifiication in last para. Also "Srongly nonlocal", does not convey what was proved, namely no gate needs more than the resource for $S$. Also appended locals in the $S$ on the right also. }

The minimum value of gate-typicality is reached only for local gates,
$g_t(U^A \otimes U^B)=0$, 
while the mean value averaged over the Haar measure reads $\ovl{g_{t}(U)}^U=1$. 
The distribution $P(g_t)$ obtained for the Haar random unitaries
%as $U$ varies over the $\cue_{N^2}$ is 
is symmetric with respect to its average value. 
Since for large $N$ this distribution is strongly 
 concentrated at the mean value $\ovl{g_{t}}=1$, 
it is appropriate to call $g_t$ as the {\sl gate-typicality}.
%. This quantity is referred to below as ``gate-typicality" with the understanding that 
Note that large deviations, including gates close to $S$ with  
$g_t \approx 2$, and nearly local gates with 
% gates near the {\sc swap})) and small values (
$g_t \approx 0$,
%  gates near product) of this measure 
are rare and atypical. 
Thus for any unitary gate $U$
 the quantity $|g_t(U)-1|$ is a measure
of its non-typicality.

% atypical while values around $1$ are typical.

\noindent {\it Effect of one intermediate local operation:}

\begin{figure}[htp!]
  \centering
 \includegraphics[scale=0.5,angle=0]{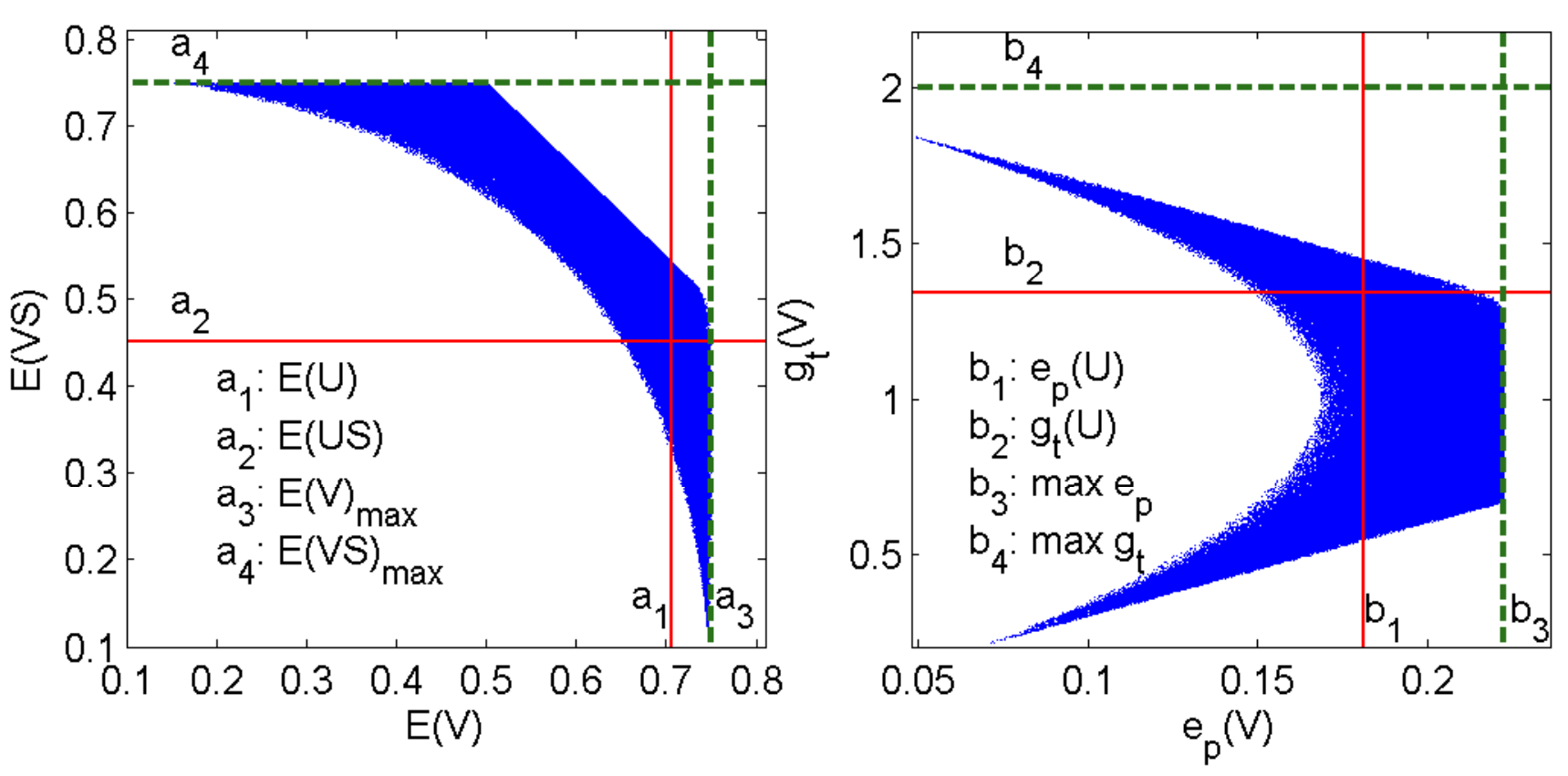}
\caption{Operator entanglements $E(V)$ {\it vs} $E(VS)$ (left), and entangling power {\it vs} gate-typicality (right) for two qubits.
Here $U$ is a fixed random entangling gate, 
$V=\sqrt{U} (U^A \otimes U^B) \sqrt{U}$ where local gates $U^A$,  $U^B$ are sampled 
randomly according to the Haar measure on $U(2)$.
%uniformly from the qubit subspaces, 
Lines (with labels indicating their meaning)
illustrate that local gates can 
increase both the entangling power and the gate-typicality of $U$. }
 \label{fig:epenl}
  \end{figure}

Let $V=\sqrt{U} (U^A \otimes U^B) \sqrt{U}$,
where %(The Circular Unitary Ensemble, $\cue_N$, with $N=4$), and the 
one qubit local unitaries  $U^A$ and $U^B$ are Haar random unitaries from $U(2)$.
Fig.~\ref{fig:epenl} shows the pairs $\{E(V), E(VS)\}$ and $\{e_p(V), g_{t}(V)\}$ for a fixed 
two--qubit gate $U$ picked at random according to the Haar measure on $U(4)$. % the space of $4\times 4$ unitary matrices 
Here 
%matricelocal unitaries are one qubit gates sampled from the $\cue_2$. 
The operator $\sqrt{U}$ has the same eigenvectors as $U$, and its eigenvalues are $e^{i\phi/2}$, where the eigenvalues of $U$ are $e^{i \phi}$ and $-\pi < \phi \le \pi$. 

It is clear from Fig.~\ref{fig:epenl} that there exists a nonzero measure of local operators that can enhance the entangling power and gate-typicality of $U$, indicated by the solid lines.
% and some of them increase both quantities. 
The same holds for the operator entanglements $E(V)$ and $E(VS)$ as demonstrated in Fig. 1.
%
% and it is interesting that except for $g_{t}$, there exist local operators
% such that the other measures approach the maximum values.
%%  KAROL %% but this feature
%% may depend on the choice of $U$ and can be gate specific => thus I comment it out...
%
The mean entangling power, averaged over the Haar measure on $U(N^2)$ reads
% $e_p(U)$ when $U$ is sampled from the $\text{CUE}_{N^2}$ is 
$\ovl{e_p}=(N-1)^2/(N^2+1)$ \cite{Zanardi2000, Zanardi2001}. 
A sufficient condition for the existence of local operators 
which can increase $e_p$ and $g_{t}$ follows as a corollary to 
the main theorem formulated below.
\medskip 

%However $E(U)$ can reach the maximum value ($=E(S)$) for gates that differ from the {\sc swap} not just by products of local operators. For example for two qubits, $E(\text{\sc cnot} \,S)=3/4=E(S)$, but $g_{t}(\text{\sc cnot}\, S)=2/3<g_{t}(S)=2$. Furthermore, unlike $e_p(U)$, $g_{t}(U)$ distinguishes whether the interlab (linear entropy) entanglement in the split $AA'|BB'$ (see Fig.~\ref{fig:ancilla}) is more than that of the cross split $AB'|A'B$ or not. Iff this is case $E(U)>E(US)$ and $g_{t}(U)>1$, and the operator may be characterized as being swap-like, else it is product-like. This is reinforced by the observation that $g_{t}(U) \lessgtr 1$ implies  $g_{t}(US) \gtrless 1$, while $e_p(U)=e_p(US)$.

 {\it Multiple iterations and averaging over local unitaries:} Local gates,
 although they have no entangling power themselves, catalyze the entangling power and
 gate-typicality when interlaced between multiple %(say $n$) 
 uses of $U$.
% The interaction strength of the interlaced gate $[U(U^A \otimes U^B)]^n$
%can be significantly larger than this achieved by $U^n$ alone.
An exact statement concerning the strength averaged over local gates 
 is stated now as the central result of this work.
\begin{theorem}
\label{thm:1}
Consider $n$ identical nonlocal operators $U$
 interlaced by local gates $W_j=U^A_j \otimes U^B_j$:
%\beq
%\label{eqn:Un}
% U^{(n)}\equiv U \, U^A_{n-1}\otimes U^B_{n-1} \, U \cdots U^A_1 \otimes U^B_1 \, U, \, n \ge 1,\, U^{(1)} \equiv U.
%\eeq
\beq
\label{eqn:Un}
 U^{(n)}\equiv U \, W_{n-1} \, U \cdots W_1 \, U, \ \  n \ge 1,\ \  U^{(1)} \equiv U.
\eeq
Then the mean entangling power and the mean gate-typicality read
%averaging over the local $\cue_N$ operators, then 
%\beq
%\begin{split}
%\begin{gather}
\begin{align}
\label{eqn:epxi}
\br e_p(U^{(n)}) \kt_W=\ovl{e_p}\left[ 1-\left(1-\dfrac{e_p(U)}{\ovl{e_p}}\right)^n \right],\\
\label{eqn:epeta}
\text{and} \ \ \br g_{t}(U^{(n)}) \kt_W= 1-[1-g_{t}(U)]^n , 
%\end{split}
%\end{gather}
\end{align}
%\eeq
where $\br \; \kt _W$ denotes the average with respect to the Haar measure on $U(N)$.
\end{theorem}
% Optimal $U$ defined for $e_p(U)$ are such that $E(U)=E(US)=E(S)$ when the entangling power is the maximum possible \cite{zanardi00}. However in this case $g_{t}(U)=1$, which is just the average operator nonlocality. In fact it is easy to see that for $g_{t}(U)>1$, $e_p(U)<(N-1)/(N+1)$, the maximum possible entangling power. The extreme case, reached only when $U$ is the {\sc swap}, is when $g_{t}(U)=2$ and the entangling power vanishes. 

% While $e_p(U)$ measures entangling power when the unitary operator is allowed to act on product states of $A$ and $B$, $E(U)$ can be interpreted as the entanglement generated between the $A$ and $B$ laboratories that include ancilla. The contrast is seen starkly with the {\sc swap} operator as $e_p(\text{\sc swap})=0$ while $E(\text{\sc swap})=1-1/N^2$, the maximum possible. 
\noindent {\bf Proof:} Define the mean purities 
of states generated from $U^{(n)}$ of Eq.~\eqref{eqn:Un} 
and averaged over local unitaries $\{U_1^A\cdots U_{n-1}^B\}$ as 
\beq
\label{eqn:XnYn}
X_n = \br \tr[\rho_R^2(U^{(n)}]\kt_W ,\  \  Y_n = \br \tr[\rho_{T}^2(U^{(n)})]\kt_W,
\eeq
so that $\br E(U^{(n)}) \kt_W=1-X_n$,  $\br E(S U^{(n)}) \kt_W=1-Y_n$, and $X_1=\tr[\rho_{R}^2(U)], \ Y_1=\tr[\rho_{T}^2(U)]$.
Given a nonlocal operator $U$, and hence $(X_1, \,Y_1)$, a linear affine 
% (a linear affine transformation), 
iterative scheme $(X_n,Y_n) \mapsto (X_{n+1},Y_{n+1})$, 
follows from the independence of the local unitaries 
-- see Supplementary Material for the details,  
\begin{widetext}
\beq
\label{eqn:rec}
\begin{split}
X_{n+1}&=\frac{1}{(N^2-1)^2}\left[ 2(N^2+1) - N^2 (2X_1+2Y_1+2X_n +2Y_n -X_1 Y_n-Y_1 X_n) + N^4 (Y_1 Y_n +X_1 X_n) \right],\\
Y_{n+1}&=\frac{1}{(N^2-1)^2}\left[ 2(N^2+1) - N^2 (2X_1+2Y_1+2X_n +2Y_n -Y_1 Y_n -X_1 X_n) + N^4 ( X_1 Y_n+Y_1 X_n ) \right].
\end{split}
\eeq
\end{widetext}
It is clear that the combinations $X_n  \pm Y_n$ do separate
so these quantities are convenient to iterate. Thus defining
\beq
\label{eqn:xieta}
\xi_n = C_N\left(X_n +Y_n -\frac{4}{N^2+1}\right), \; \eta_n =D_N(X_n-Y_n),
\eeq
where $C_N= N^2(N^2+1)/(N^2-1)^2$, $D_N=N^2/(N^2-1)$ leads to simple recursion relations,
$\xi_{n+1}= \xi_1 \xi_{n}, \;\eta_{n+1}=\eta_1 \eta_{n}$, with solutions $\xi_{n}= \xi_1^n, \;\eta_{n}= \eta_1^{n}$.
It is easy to generalize this reasoning for the case in which  nonlocal operators are 
different at each iteration in Eq.~\eqref{eqn:Un}. Denoting  different values of $\xi_1$ 
as $\xi_{1k}$, then $\xi_n=\prod_{k=1}^n \xi_{1k}$. Note that  $\xi_n$ 
is averaged over the local unitaries, while  $\xi_{1k}$ are derived from the purities of the corresponding density matrices as in Eq.~\eqref{eqn:rhoRT}.

The quantity $\xi_1$ is related to the entangling power of $U$ 
and, remarkably, 
%on using the definitions 
it follows from the definition 
that $\xi_1=1-e_p(U)/\overline{e_p}$. This in turn, 
along with $\xi_n =\xi_1^n$, results in  Eq.~\eqref{eqn:epxi}.
%This results in 
%\beq
%\br e_p(U^{(n)}) \kt=\ovl{e_p}\, (1-\xi_n)=\ovl{e_p}\, (1-\xi_1^n),
%\eeq 
%and hence Eq.~\eqref{eqn:epxi} follows. 
The complementary quantity $\eta_n$ distinguishes between the {\sc swap} and local gates, and 
provides additional motivation to introduce the gate-typicality  \eqref{eqn:enlU}. 
Hence using the definition \eqref{eqn:xieta} of $\eta_1$ one can show that 
%it is verified that
$\eta_1=1-g_{t}(U)$ and the advertised exponential convergence \eqref{eqn:epeta} follows.
$\square$
\smallskip

 \begin{corollary}
\label{corr:1}
If $e_p(U)< e_p(\sqrt{U})\, (2-e_p(\sqrt{U})/\ovl{e_p})$, then there exist local operators $U^A$ and $U^B$ 
such that $e_p(U) < e_p(\sqrt{U} (U^A \otimes U^B) \sqrt{U})$.
\end{corollary}

The corollary follows as the theorem implies that if $\br e_p(U^{(2)})\kt_W>e_p(U^2)$ then $e_p(U^2)<e_p(U)(2-e_p(U)/\overline{e_p})$. 
%In this case the local unitary averaged 
The entangling power of $UWU$ averaged over local unitaries
is larger than the entangling power of $U^2$ and therefore there exist
 members of the ensemble of local unitary operators such that 
$e_p(U (U^A \otimes U^B) U)>e_p(U^2)$. 
If the nonlocal gate $U$ is chosen at random from $U(4)$
% $\cue_4$, 
numerical results indicate that about 28$\%$ of them satisfy the condition in the corollary, one such realization is shown in Fig.~\ref{fig:epenl}. 
 Thus as long as $e_p(U)\ne 0$, $\br e_p(U^{(n)}) \kt_W \rightarrow \ovl{e_p}$
 as $n \rightarrow \infty$, and the convergence is exponential with the rate which 
 depends on the entangling power $e_p(U)$.
%of the nonlocal unitary.  
A similar statement about the gate-typicality $g_t$  follows. 
Thus if $g_{t}(U)<g_{t}(\sqrt{U})\, (2-g_{t}(\sqrt{U}))$, then there exist local operators such that 
$ g_{t}(U) < g_{t}(\sqrt{U} (U^A \otimes U^B) \sqrt{U})$.

Note that the above statements solve completely and exactly 
the problem of finding the 
entangling power and gate-typicality of an iterated nonlocal operation 
averaged over random local gates which interlace the dynamics.
%interlaced with independent local unitaries. 
Two convergence rates follow, $\log|1/\xi_1|$ for the 
entangling power and $\log|1/\eta_1|$ for the gate-typicality. 
Any entangling gate $U$ iterated with interlaced random local unitaries will 
lead to typical entangling power and mean typicality % $g_t(U)$ 
at these rates. 
Also the purities tend to their %$\cue_{N^2}$ 
mean values:
\beq
X_{\infty}= Y_{\infty}=\frac{2}{N^2+1},
\eeq
is the fixed point {\it independent} of unitary $U$, as long as 
it is not itself a local operator or the swap gate.

The ranges of $\xi_1$ and $\eta_1$ are
\beq
-\dfrac{2}{(N^2-1)} \le \xi_1 \le 1, \ \ \ \  -1 \le \eta_1 \le 1.
\eeq
The upper-bounds are reached by unitary operators that are local. In addition $\xi_1=1$ also for the {\sc swap} gate, while for $\eta_1$ the lower-bound of $-1$ is reached {\it only} in this case. The lower-bound of $\xi_1$ follows from fact that $X_1$ and $Y_1$ are density matrix purities and hence cannot be less than $1/N^2$ each. However,
this is not a tight bound for $N=2$ and this is related to the nonexistence of absolutely maximally entangled states of four qubits \cite{Higuchi2000} 
and of multiunitary matrices of order four \cite{Goyeneche2015}. 
The bound is tight in all dimensions except  $2$ and possibly $6$,
which follows also from a previous study of the entangling power of permutations \cite{Clarisse2005}. 
As shown below for $N=2$ the minimum value of $\xi_1$ is achieved if 
$U$ is the {\sc cnot} gate.

\medskip 

\noindent {\it Examples:} % Of the many possible nonlocal operations, three are considered, 
To present our Theorem in action we now discuss three  paradigmatic unitary gates:
(a) the two qubit \textsc{cnot} gate, 
(b) unitaries $U \equiv U_d$ consisting of only diagonal elements, and 
(c)  higher dimensional controlled gates. 

\noindent (a) Two qubit \textsc{cnot} gate reads
 $|0\kt \br0| \otimes \mathbb{1}+|1\kt \br 1| \otimes \sigma_x$, and simple calculations 
%based on their definitions 
yield
% $\rho_R(\textsc{cnot})=(|00\kt \br 00|+|11\kt \br 11|)/2$, and $\rho_T(\textsc{cnot})=I_4/4$, where $I_4$ is the identity matrix of dimension $4$, and hence 
$X_1=1/2, \, Y_1=1/4$,  $e_p(\textsc{cnot})=2/9$ and $g_{t}(\textsc{cnot})=2/3$. Using Eqs.~(\ref{eqn:epxi},\ref{eqn:epeta}) results in $\br e_p(\textsc{cnot}^{(n)}) \kt_W=$
\beq
%\begin{split}
\frac{1}{5}\left(1-\frac{(-1)^n}{9^n}\right),\, \br g_{t}(\textsc{cnot}^{(n)}) \kt_W= 1-\frac{1}{3^n}.
%\end{split}
%X_n=\dfrac{2}{5} +\dfrac{1}{8} \dfrac{1}{3^{n-1}}+\dfrac{(-1)^n}{40} \dfrac{1}{9^{n-1}},
\eeq
The entangling power decreases from the maximum possible $2/9$ 
%\textcolor{red}{(The maximum possible is $\frac{N-1}{N+1}$ which is 1/3 for two qubit gate!)}
to the average $1/5$ at the rate of $2 \log 3$, while the gate-typicality increases from $2/3$ to $1$ at the rate of $\log 3$.

%With $n=4$ applications of \textsc{cnot} the expected purity of $\rho_R$ is $295/729\approx 0.4046$, and the average of the measure $E(\textsc{cnot}_4)$  $\approx 0.595$ is very close to the CUE average of $0.6.$ 

\noindent (b) Random diagonal unitaries $U_d$ studied in 
 \cite{Nakata2012,Lakshminarayan2014} 
 arise as interactions in many Floquet models.
Note first that the square of a diagonal unitary matrix remains diagonal, 
so the entangling power and the gate-typicality of $U_d^n$ 
remains approximately the same during the time evolution.
However, if the evolution is interlaced by local dynamics
the situation changes dramatically.
Applying Eq. (\ref{eqn:XnYn}) we obtain  
\beq
\label{eqn:X1avg}
\overline{X_1}=\overline{\tr (\rho_R^2(U_d))}^{U_d} = \dfrac{2N-1}{N^2},\;  \text{var}(X_1)=2 \dfrac{(N-1)^2}{N^6}.
\eeq
where the bar indicates additional averaging over the diagonal elements, which are uniform random phases --
the average  $\overline{X_1}$ over the unimodular ensemble is derived
% the unimodular ensemble of
in  \cite{Lakshminarayan2014}.
 As the gate $U_d$ is diagonal it is invariant with respect to partial transposition, so 
 $\rho_{T}(U_d)=I_{N^2}/N^2$ and $Y_1=1/N^2$. 
 Thus for generic diagonal unitaries one has 
  $X_1 \sim 2/N$ and $Y_1=1/N^2$. These values imply the following behavior 
%using the solution to the iteration leads 
for $n \ll N$,  
%\beq
%\label{eqn:XYdiag}
%\begin{split}
%X_n-X_{\infty} &= \dfrac{2^n}{N^n}  \left[ 1+ \mathcal{O}\left(\frac{n}{N}\right)\right], \\ Y_n-Y_{\infty}&=\mathcal{O}\left(\frac{2^n n}{N^{n+1}}\right).
%\end{split}
%\eeq
\beq
\label{eqn:XYdiag}
\Delta X_n= \dfrac{2^n}{N^n}  \left[ 1+ \mathcal{O}\left(\frac{n}{N}\right)\right], \Delta Y_n=\mathcal{O}\left(\frac{2^n n}{N^{n+1}}\right),
\eeq
where $\Delta X_n =X_n -X_{\infty}$ and $\Delta Y_n =Y_n -Y_{\infty}$.
Thus  $Y_n$  almost reaches its typical value only after two generic
diagonal gates %nonlocal operators 
as $Y_2 \sim 2/N^2$. 
%Much closer to $X_{\infty}$ than was $X_1\sim 1/N$, $X_2$ still goes as  $6/N^2$, and  reflects significant deviations from the typical. 
The value of  $X_2 \sim 6/N^2$ and  reflects significant deviations from stationarity after two iterations, while $X_3 \sim 2/N^2$, the same leading order as $X_{\infty}$.  

\begin{figure}[h]
  \centering
\includegraphics[width=0.35\textwidth]{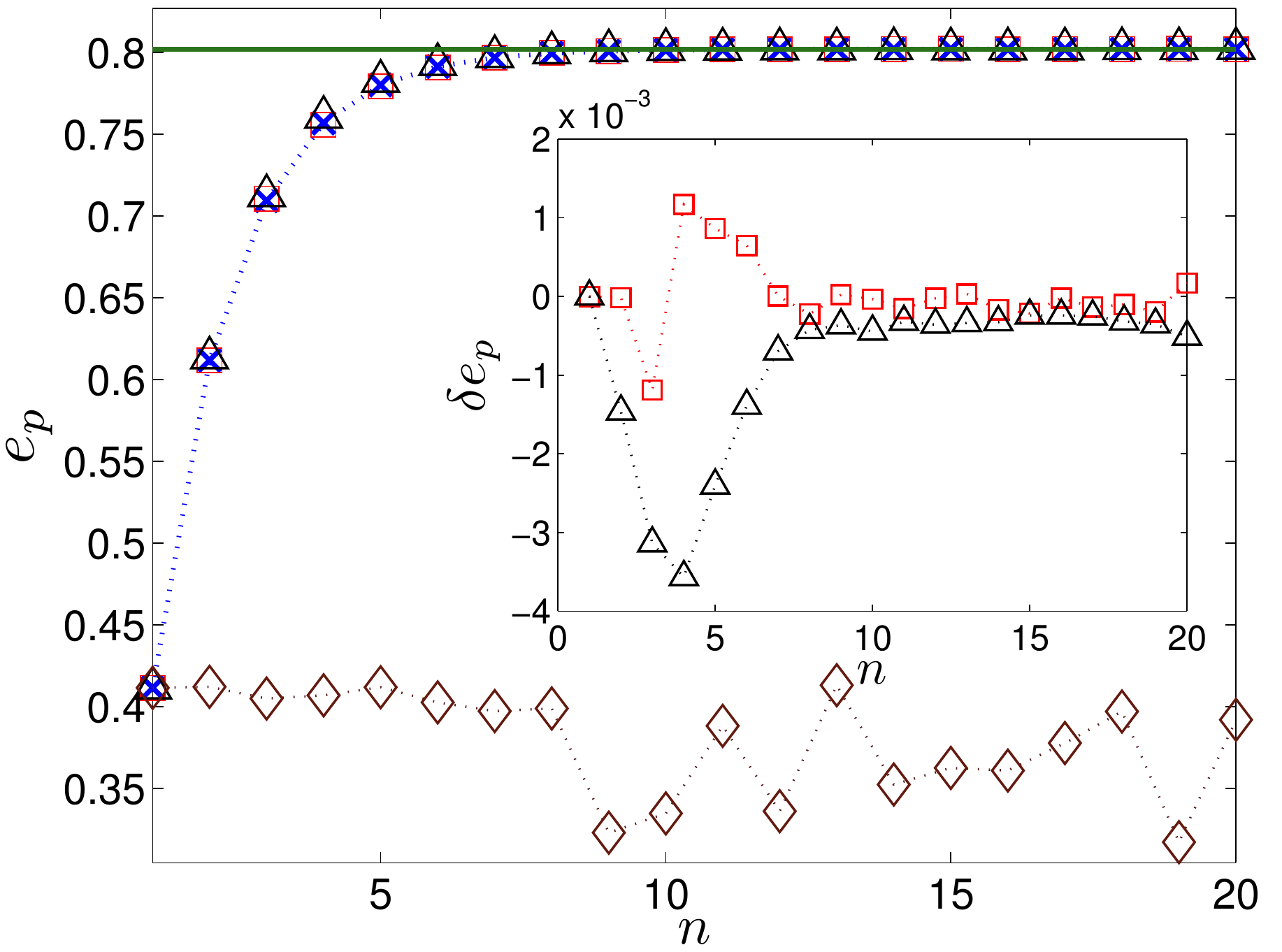}
  \caption{Entangling power $e_p$ after $n$ actions of a controlled unitary gate $U \in U(N^2)$ 
plotted for $N=10$.
Data from one realization of $U^{(n)}$ (\textcolor{red}{$\square$}) and one realization of $[(U^A\otimes U^B)\,U]^n$ ($\triangle$) are shown.  Values averaged over local gates according to  Eq.~\eqref{eqn:epxi} are indicated by (\textcolor{blue}{$\times$}),
 while the insets show deviations from this average.
Horizontal line denotes the average over the unitary group.
The case $U^n$ with no interlacing gates, for which entangling power do not converge 
to typical values, is marked by (\textcolor{Brown}{$\Diamond$}).}
 \label{fig:Xnvsn}
\end{figure}

These results imply that, up to the lowest order in $\nicefrac{1}{N}$, the entangling power increases from the value
$1-(\nicefrac{4}{N})$, typical for diagonal unitaries  to the mean value over the Haar measure, 
  $\overline{e_p} = 1-(\nicefrac{2}{N})$.
On the other hand, the gate-typicality $g_{t}$ increases 
from the value $1-(\nicefrac{2}{N})$ characteristic to diagonal unitaries to $1$ under the influence of 
random local gates. 

%A more elaborate computation yield unimodular ensemble averages
%\beq
%\overline{X_2}=\dfrac{6}{(N+1)^2},\, 
%\overline{Y_2}=\dfrac{2 (N^4+N^2+1)}{N^4(N+1)^2},
%\eeq
%consistent with the above $n=2$ case.

% This is reflected in the eigenvalues and singular values of $(U_d\, U^A_1\otimes U^B_1 \,U_d)_{R,T}$ which is treated as an ensemble consisting of different local unitaries and different diagonal nonlocal operators. Using the recursion in Eq.~\eqref{eqn:rec} and the third moment $\overline{X_1^3}$, $\overline{X_3}\sim 2/N^2$ which indicates that 3 applications of $U_d$ is required to reach the CUE average.
%
\noindent (c) Controlled unitaries can be implemented using a simple nonlocal protocol \cite{Yu2010} with prior entanglement. 
%Removed 2 references cohen13l,chen14schmidt}
%It is shown that any bipartite unitary of Schmidt rank $2$ and $3$ is locally equivalent to a controlled unitary\cite{cohen13l,chen14schmidt}.
 Consider a controlled gate $U=P_1^A\otimes\mathbb{1}^B+P_2^A\otimes V^B $,
%% Change   U^B -- > V^B to avoid confusion with local evolution
%%  Karol
 where $P_1^A+P_2^A=\mathbb{1}_{N}$ and $P_i^AP_j^A=\delta_{ij}P_i^A$, 
 and  $V^B$ is some $N$ dimensional unitary operator.
In this case $X_1= (N^2+|{\tr \,V^B}|^2)/2N^2$, $Y_1=1/N^2$. 
For $n \ll N$  the iteration results in 
\beq
\label{eqn:XYcU}
\begin{split}
\Delta X_n &= \dfrac{1}{2^n}\left[1+ \mathcal{O}\left(\dfrac{n}{N^2}\right)\right], \\ \Delta Y_n&= \dfrac{1}{2^n\, N^2}\left[-(n+1)+ \mathcal{O}\left(\dfrac{n}{N^2}\right)\right].
\end{split}
\eeq
The details of the gate $V^B$ are relevant by its trace
only up to higher orders represented by the symbol $\mathcal{O}$.
The contrast with Eq.~\eqref{eqn:XYdiag} is apparent as this indicates a much slower 
convergence to the asymptotic values. 
It is also clear that the quantity $Y_n$ approaches its limiting value faster than $X_n$. 
For instance, for $n=2$ one has $X_2\sim 1/4$, while $Y_2\sim 5/(4 N^2)$. The Haar 
average is reached within precision  $\mathcal{O}(1/N^4)$  by $X_n$ for time
$n\sim 4 \log_2N$.

Under the action of random local gates
the entangling power increases  % (up to order $1/N^2$)
from the initial value close to $\nicefrac{1}{2}-\nicefrac{1}{N}$ to the stationary value of $1-\nicefrac{2}{N}$,
while the gate-typicality increases from $\nicefrac{1}{2}$ to $1$.
If no local operators were used, observe first that $U^n$ is also a controlled unitary.
Assuming now that $V^B$ is taken as a Haar random unitary from $U(N)$ % typical $\text{CUE}_N$ matrix, 
then the average form factor  $\br |\tr (V^B)^{n}|^2\kt$ 
equal to $n$ for $n\le N$ and to  $N$ for $n>N$ -- see  \cite{Haake1996}  --
implies that the entangling power
of $U^n$  decreases to $\nicefrac{1}{2}-(\nicefrac{3}{2N})$ for times long enough,
 while the gate-typicality decreases to $\nicefrac{1}{2}-(\nicefrac{1}{2N})$.

%One may also construct an ensemble of random controlled gates by sampling $U^B$ according to the CUE$_N$. For example detailed calculations involving the form factors of the CUE~\cite{haake13} give the additionally averaged quantities
%\beq
%\overline{X_2}=\frac{N^6+2N^4-6N^2+4}{4N^2(N^2-1)^2},\,
%\overline{Y_2}=\frac{5N^4 - 10N^2 +6}{4N^2(N^2-1)^2},
%\eeq           
%consistent with the unaveraged $n=2$ cases in Eq.~\eqref{eqn:XYcU}.
%\section*{Acknowledgments}

Simulations in Fig.~\ref{fig:Xnvsn} illustrates that the formulae
derived for entangling power  averaged over an ensemble of local interlacing gates form 
excellent approximations to time evolution even for 
a {\it single} realization of local gates.
% and the exponential approach to typicality is observed even here. 
%%
More remarkably, formulae derived also work if 
the {\it same} local gates are applied at every iteration,
 so that $U^{(n)}=(U^A \otimes U^B\, U)^n$ -- see inset for the smallness of the deviations.
Thus these results are of direct relevance to the study of iterated coupled quantum Floquet systems such as in \cite{Lakshminarayan01,Dobrzanski2004}. 
Similar qualitative behavior is observed for gate-typicality as well. 
% it still remains a very good approximation, as the inset illustrates. 
Thus correlations introduced by the repeated action of local gates are not significant, and
% if they are typical. 
%The controlled unitary with a random, but fixed, $V^B$ is used in these calculations. 
%Results presented for a fixed controlled unitary of order $N=10$
%demonstrate that 
the entangling power and gate-typicality continue to reach their asymptotic values
exponentially fast. 
%Similar results have been obtained for the diagonal operators. 

%\begin{figure}[h!]
%  \centering
%  \includegraphics[width=.45\textwidth]{figures/eigval_singval_diag_reshuf_50.eps}
%\caption{The eigenvalues of the realigned $U^{(n)}_R$
% when $U$ is a diagonal random  matrix (top row) where $N=50$, while the middle row shows their radial density. The third row shows the distribution of the eigenvalues of  $N^2\rho_R(U^{(n)})$ or squared singular values of $U^{(n)}_R$, with the solid curve in the final figure is the Marcenko-Pastur distribution. }
%\label{fig:DiagMP}
%\end{figure}
%\begin{figure}[h!]
%  \centering
%\includegraphics[width=.45\textwidth]{figures/eigval_singval_cu_reshuf_50.eps}
%  \caption{Same as the previous figure, but when $U$ is a controlled random unitary matrix. The figure illustrates the relatively longer time scale it takes before typicality sets in.}
%  \label{fig:CUMP}
%\end{figure}
%
\medskip 

\noindent{\it Summary and outlook:} Iterating nonlocal unitary operators with interlaced local dynamics is a typical scenario in 
both time evolution and simple quantum circuits. 
We have shown here that two quantities characterizing the interaction strength, namely
the entangling power and gate-typicality are significantly  modified  by subsequent application of local gates. We have shown that both quantities converge exponentially to their
asymptotic values and computed the mean
convergence rates under the assumption that local gates are 
distributed randomly according to the Haar measure on $U(N)$.
%
%by the local operations and as illustrated in two examples both may increase towards typical %values. This approach to typical values is exponential 
% and reminiscent of approach to equilibrium. 

As typical for ergodic problems, 
a generic realization is shown to closely follow the average behavior. 
Our analytic predictions hold even when
%continue to describe well the time evolution in which 
%Even in the case when 
the same local unitary gate is applied several times. 
%this continues to be a good approximation.
Additional numerical investigations show that
other moments $\tr(\rho^k_R(U^{(n)})$ and $\tr(\rho^k_T(U^{(n)}))$, with $k \ne 2$,
and the von Neumann entropies 
also exponentially approach their limiting values,
% CUE$_{N^2}$ value 
%is indicated by numerical results (not displayed) 
%which show that 
as the density of the rescaled eigenvalues of 
$\rho_{R}(U^{(n)})$ and $\rho_{T}(U^{(n)})$ 
for large $N$ approaches  the Marcenko-Pastur distribution \cite{Marchenko1967}.

A detailed study and further interpretation of the gate-typicality is called for.
It is important to investigate the extent to which nonrandom local operators
influence the approach to equilibrium of a periodically interlaced
unitary dynamics. Applications to Floquet models of condensed matter physics 
and quantum chaos would be interesting. Generalizations to
multipartite settings as well as to generalized quantum operations 
%while not straightforward, 
are worth studying.

\medskip 

\acknowledgements{We are grateful to Som Bandyopadhyay for discussions on the ancilla interpretation and Steven Tomsovic for comments.
This work was supported by the Polish National Science Center under the project number DEC-2015/18/A/ST2/00274, by the John Templeton Foundation under the project No. 56033, and the Indian DST (INSPIRE) project PHY1415305DSTXPRAN. }

%\textcolor{red}{(We din't define $n^*$, we are using $n_k$)}. 
%In fact the spectra of the non-unitary operators $U^{(n)}_{R}$ and $U^{(n)}_{T}$ (realigned and partial transposed $U^{(n)}$ whose singular values are the proportional to the density matrix operators), are also interesting and reflect a tendency to get increasingly uniformly distributed on the unit disk (the so-called universal Girko circle law \cite{}) along with the approach to typicality and the realization of the Marcenko-Pastur distribution. The Figs.~(\ref{fig:DiagMP},\ref{fig:CUMP}) display numerical results for the realigned operators in the case of diagonal unitaries and controlled unitary operators. The rapidity with which typicality is reached in the case of diagonal nonlocal operators in comparison to the controlled unitary is also reflected in these.

\bibliography{ent_local}
%\bibliography{quantummodify,rmtmodify,classicalchaos}

\newpage

\appendix
\begin{center}
{\bf \large Supplementary Material}
\end{center}
\noindent{\it A. Operator entanglement and the ancilla interpretation:}

\begin{figure}[htp!]
  \centering
  \includegraphics[scale=0.5]{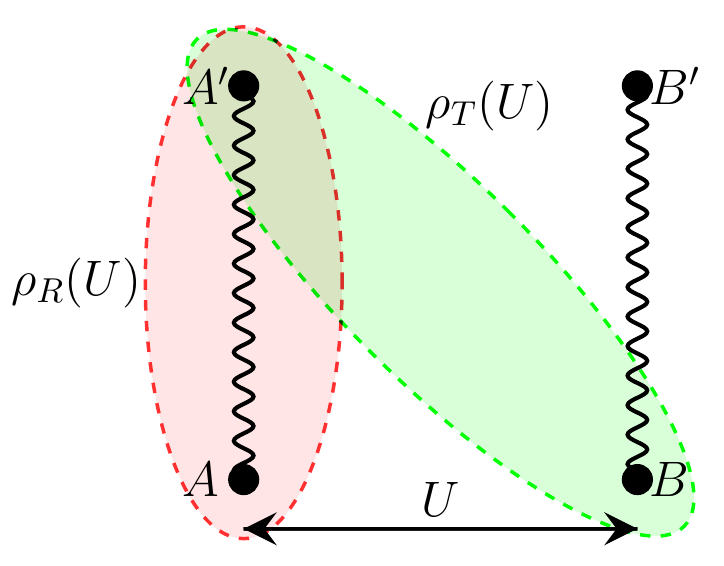}
    \caption{Operational significance of $E(U)$. It captures the entanglement generated
   by the action of $U$ on a pure state of systems $A,B$ along with ancillas 
   $A'$ and $B'$, across the splitting $AA'|BB'$.  
  Similarly $E(US)$ is the entanglement with respect to the partition
  $AB'|A'B$.}
    \label{fig:ancilla}
  \end{figure}

It is useful to view $E(U)$ as entanglement in a pure state
between $A$ and $B$ along with a bi-partite ancilla $A'B'$ -- see Fig.~\ref{fig:ancilla}. 
Let  $AA'$ be in the  standard maximally entangled state $|\phi_{AA'}^+\kt=\sum_{j=1}^N|jj\kt/\sqrt{N}$, 
with $A'$ being an ancilla with $A$, also of dimension $N$, 
and let $BB'$ be in a similar state.
 If $U$ acts between  $A$ and $B$ subsystems,
 the reduced density matrices of  $AA'$ and $A'B$ are respectively,
$\rho_{R}(U)= U_RU_R^{\dagger}/N^2$, and $ \rho_{T}(U)=S \,U_{T}U_{T}^{\dagger}S/N^2$.
Here $S$ is the {\sc swap} operator, $U_R$ is the reshuffling of $U$,
 while $U_{T}$ is its partial transpose with respect to $A$ and are defined in the main text. 
The operator linear entropy $E(U)$ is thus the linear entropy of  the state $\rho_{R}(U)$ 
and measures the entanglement in this quadripartite state 
with respect to the partition $AA'|BB'$.
 Since  $E(SU)=E(US)$ it represents the entanglement of the same state with respect to 
the partition  $AB'|A'B$ partition  -- see Fig.  \ref{fig:ancilla}.

{\it B. Proof for the iterative scheme:}
% The approach is as follows: Consider the nonlocal unitaries $U$ and $V$ and the local operator $U^A_1\otimes U^B_1$ on the bipartite Hilbert space $\mathcal{H}^N\otimes\mathcal{H}^N$. Using $X_1$ and $Y_1$  i.e. $\tr(\rho_{R,T}^2)$ of $U$ and $V$, $X_2$ and $Y_2$ of $U\,U^A_1\otimes U^B_1\,V$ can be found for which the proof is given below. With $X_2$, $Y_2$ and by realizing $V=U\,U^A_2\otimes U^B_2\,U\ldots U^A_{n-1}\otimes U^B_{n-1}\, U$, an iterative method can be used to find $X_n$, $Y_n$ given in Eq.~\eqref{eqn:rec}. For example if $V = U\,U^A_2\otimes U^B_2\,U\, U^A_{3}\otimes U^B_{3}\, U$, $X_3$ and $Y_3$ can be calculated using $X_1$, $Y_1$ and $X_2$, $Y_2$.

Consider the operator $\mathcal{N}=U\, (U^A\otimes U^B)\,V$. Denoting the matrix elements $\br ij|U|pq\kt\equiv U_{\substack{ij\\pq}}$ and summing over repeated indices 
one arrives at 
\begin{widetext}
\beq
\begin{split}
X_{UV} &\equiv\br \tr [\rho_R^2(\mathcal{N})]\kt _{U^A,U^B}= \frac{1}{N^4}U_{\substack{k\alpha\\ m_1\gamma_1}}V_{\substack{m_2\gamma_2\\ l\beta}}U_{\substack{\alpha^{\prime}k^{\prime}\\ m_3\gamma_3}}V_{\substack{m_4\gamma_4\\ \beta^{\prime}l^{\prime}}}\overline{U}_{\substack{\alpha^{\prime}\alpha\\ m_5\gamma_5}}\overline{V}_{\substack{m_6\gamma_6\\ \beta^{\prime}\beta}}\overline{U}_{\substack{kk^{\prime}\\ m_7\gamma_7}}\overline{V}_{\substack{m_8\gamma_8\\ ll^{\prime}}} \left \br \left( \mathcal{U}^A\overline{\mathcal{U}}^A\right)_{[m]} \right \kt \left \br\left( \mathcal{U}^B\overline{\mathcal{U}}^B\right)_{[\gamma]}\right \kt,\\
\text{where} \;\; &\left( \mathcal{U}^A\overline{\mathcal{U}}^A\right)_{[m]} = U^A_{m_1m_2}U^A_{m_3m_4}\overline{U}^A_{m_5m_6}\overline{U}^A_{m_7m_8};\quad\left( \mathcal{U}^B\overline{\mathcal{U}}^B\right)_{[\gamma]} = U^B_{\gamma_1\gamma_2}U^B_{\gamma_3\gamma_4}\overline{U}^B_{\gamma_5\gamma_6}\overline{U}^B_{\gamma_7\gamma_8},
\end{split}
\eeq
\end{widetext}
%\beq
%\begin{split}
%X_2=&\frac{1}{N^4}U_{\substack{k\alpha\\ m_1\gamma_1}}V_{\substack{m_2\gamma_2\\ l\beta}}U_{\substack{\alpha^{\prime}k^{\prime}\\ m_3\gamma_3}}V_{\substack{m_4\gamma_4\\ \beta^{\prime}l^{\prime}}}\overline{U}_{\substack{\alpha^{\prime}\alpha\\ m_5\gamma_5}}\overline{V}_{\substack{m_6\gamma_6\\ \beta^{\prime}\beta}}\overline{U}_{\substack{kk^{\prime}\\ m_7\gamma_7}}\overline{V}_{\substack{m_8\gamma_8\\ ll^{\prime}}}\\
 %  & \br U^A_{m_1m_2}U^A_{m_3m_4}\overline{U}^A_{m_5m_6}\overline{U}^A_{m_7m_8}\kt\br U^B_{\gamma_1\gamma_2}U^B_{\gamma_3\gamma_4}\overline{U}^B_{\gamma_5\gamma_6}\overline{U}^B_{\gamma_7\gamma_8}\kt,
 %  \end{split}
%\eeq
and the bar indicates the complex conjugate. 
A similar expression holds for $Y_{UV}= \br \tr(\rho_{T}^2(\mathcal{N}))\kt _{U^A,U^B}$. The average over the local unitaries $U^A$, $U^B$ are  independent and such averages over the unitary group have long been known  \cite{Mello1990,Collins2006} (see \cite{Puchala2011} for a Mathematica function to calculate the averages). While they are in general expressed in terms of the so-called Weingarten functions, this particular 4-term average is simple enough:
\begin{widetext}
\beq
\label{eqn:2nd_moment}
\begin{split}
    \br U_{i_1j_1}U_{i_2j_2}\overline{U}_{i^{\prime}_1j^{\prime}_1}\overline{U}_{i^{\prime}_2j^{\prime}_2}\kt\equiv\int_{U(N)}^{} U_{i_1j_1}U_{i_2j_2}\overline{U}_{i^{\prime}_1j^{\prime}_1}\overline{U}_{i^{\prime}_2j^{\prime}_2}dU &= \frac{1}{N^2-1}(\delta_{i_1i^{\prime}_1}\delta_{i_2i^{\prime}_2}\delta_{j_1j^{\prime}_1}\delta_{j_2j^{\prime}_2} + \delta_{i_1i^{\prime}_2}\delta_{i_2i^{\prime}_1}\delta_{j_1j^{\prime}_2}\delta_{j_2j^{\prime}_1})\\
    &-\frac{1}{N(N^2-1)}(\delta_{i_1i^{\prime}_1}\delta_{i_2i^{\prime}_2}\delta_{j_1j^{\prime}_2}\delta_{j_2j^{\prime}_1} + \delta_{i_1i^{\prime}_2}\delta_{i_2i^{\prime}_1}\delta_{j_1j^{\prime}_1}\delta_{j_2j^{\prime}_2})
\end{split}
\eeq
\end{widetext}
 Thus there are 16 terms that should be computed for finding $X_2$, and a few are calculated below. Let 
\beq
\begin{split}
 \left\br\left(\mathcal{U}^A\overline{\mathcal{U}}^A\right)_{[m]} \right\kt&\equiv \frac{1}{N^2-1}[D_1+D_2 - \frac{1}{N}(D_3+D_4)],\\
\left\br\left(\mathcal{U}^B\overline{\mathcal{U}}^B\right)_{[\gamma]} \right\kt &\equiv \frac{1}{N^2-1}[D^\prime_1+D^\prime_2 - \frac{1}{N}(D^\prime_3+D^\prime_4)],
\end{split}
\eeq
where $D_i$ and $D^\prime_j$, $i$, $j=1,\ldots,4$, are products of Kronecker deltas which can be read from Eq.~\eqref{eqn:2nd_moment}.  For example, the term corresponding to $D_4D_1^\prime$ in $X_2$ is
\beq
\begin{split}
&\frac{-1}{N^5(N^2-1)^2}\delta_{m_1m_7}\delta_{m_3m_5}\delta_{m_2m_6}\delta_{m_4m_8}\delta_{\gamma_1\gamma_5}\delta_{\gamma_2\gamma_6}\delta_{\gamma_3\gamma_7}\delta_{\gamma_4\gamma_8}\\
&\qquad \quad U_{\substack{k\alpha\\ m_1\gamma_1}}V_{\substack{m_2\gamma_2\\ l\beta}}U_{\substack{\alpha^{\prime}k^{\prime}\\ m_3\gamma_3}}V_{\substack{m_4\gamma_4\\ \beta^{\prime}l^{\prime}}}\overline{U}_{\substack{\alpha^{\prime}\alpha\\ m_5\gamma_5}}\overline{V}_{\substack{m_6\gamma_6\\ \beta^{\prime}\beta}}\overline{U}_{\substack{kk^{\prime}\\ m_7\gamma_7}}\overline{V}_{\substack{m_8\gamma_8\\ ll^{\prime}}}\\
&= \frac{-1}{N^2(N^2-1)^2}\br{k\alpha\alpha^\prime k^\prime}|U\otimes U|{m_1\gamma_1m_3\gamma_3}\kt\\
&\qquad \qquad \qquad \quad \br{m_3\gamma_1 m_1\gamma_3}|U^{\dagger}\otimes U^{\dagger}|{\alpha^{\prime}\alpha kk^{\prime}}\kt\\
&= \frac{-1}{N^2(N^2-1)^2}\tr[(U\otimes U)\,S_{AA^\prime}\,(U^\dagger\otimes U^\dagger)\,S_{AA^\prime}]\\
&= \frac{-N^2}{(N^2-1)^2}\tr(\rho_{R}^2(U))\equiv \frac{-N^2}{(N^2-1)^2}X^U_1
\end{split}
\eeq
As another example, the term corresponding to $D_3D_4^\prime$ is found to be
\beq
\begin{split}
& \frac{1}{N^6(N^2-1)^2}\tr[(U\otimes U)S_{BB^\prime}(U^{\dagger}\otimes U^{\dagger})S_{AA^\prime}]\\
&\qquad \qquad \qquad\quad\tr[(V\otimes V)S_{AA^\prime}(V^{\dagger}\otimes V^{\dagger})S_{AA^\prime}]\\
&= \frac{N^2}{(N^2-1)^2}\tr(\rho_{T}^2(U))\tr(\rho_{R}^2(V))\\
&\equiv  \frac{N^2}{(N^2-1)^2}Y^U_1X^V_1.
\end{split}
\eeq
Evaluating all such terms including those for $Y_{UV}$ results in
\beq
\label{eqn:XYUV}
\begin{split} 
X_{UV}&= \frac{1}{(N^2-1)^2}\left[ 2(N^2 + 1) - 2N^2 (X^U_1+Y^U_1+X^V_1+Y^V_1)\right.\\
 &\left. +N^4(X^U_1 X^V_1+Y^U_1 Y^V_1) + N^2 (Y^U_1 X^V_1 +X^U_1 Y^V_1)\right],\\
Y_{UV}&= \frac{1}{(N^2-1)^2}\left[ 2(N^2 + 1) - 2N^2 (X^U_1+Y^U_1+X^V_1+Y^V_1)\right.\\
 &\left. +N^4(X^U_1 Y^V_1+Y^U_1 X^V_1) + N^2 (Y^U_1 Y^V_1 +X^U_1 X^V_1)\right].
    \end{split}
\eeq

Now let  $V = U\,(U^A_2\otimes U^B_2)\,U\ldots (U^A_{n-1}\otimes U^B_{n-1})\, U$, and take the average of 
both sides of Eq.~\eqref{eqn:XYUV} over all the local operators $U^A_2, \cdots U^B_{n-1}$. By definition then
$\br X_{UV} \kt _{U^A_2, \cdots U^B_{n-1}}\equiv X_n$, $X^U_1 \equiv X_1$ is independent of the 
local operators and $\br X^V_1\kt _{U^A_2, \cdots U^B_{n-1}}\equiv X_{n-1}$, with identical expressions 
for $Y$. Hence the recursion in Eq.~\eqref{eqn:rec} follows.

%\begin{comment}
%\begin{widetext}
%\beq
%\label{eqn:YUV}
%\begin{split} 
%X_2&= \frac{1}{(N^2-1)^2}\left[ 2(N^2 + 1) - 2N^2 (X^U_1+Y^U_1+X^V_1+Y^V_1)+N^4(X^U_1 Y^V_1+Y^U_1 X^V_1) + N^2 (Y^U_1 Y^V_1 +X^U_1 X^V_1)\right],\\
%Y_2&= \frac{1}{(N^2-1)^2}\left[ 2(N^2 + 1) - 2N^2 (X^U_1+Y^U_1+X^V_1+Y^V_1)+N^4(X^U_1 Y^V_1+Y^U_1 X^V_1) + N^2 (Y^U_1 Y^V_1 +X^U_1 X^V_1)\right].
%    \end{split}
%\eeq
%\end{widetext}
%A similar calculation results in
%\beq
%\label{eqn:YUV}
%\begin{split} 
%Y^{U,V}_2&= \frac{1}{(N^2-1)^2}\left[ 2(N^2 + 1) - 2N^2 (X^U_1+Y^U_1+X^V_1+Y^V_1)\right.\\
% &\qquad \left. {} +N^4(X^U_1 Y^V_1+Y^U_1 X^V_1) + N^2 (Y^U_1 Y^V_1 +X^U_1 X^V_1)\right].
%    \end{split}
%\eeq
% Replacing $V$ with $U$ in Eqs.~\eqref{eqn:XUV},\,\eqref{eqn:YUV} gives $X_2$ and $Y_2$ respectively. $X_3$ and $Y_3$ can be found by replacing $V$ with $U\,U^A_2\otimes U^B_2\,U$. In this case, $X_1^V=X_2$, $Y_1^V=Y_2$. In general, with $V=U\,U^A_2\otimes U^B_2\,W$ and by changing $W$ to a pair of different nonlocals and intermediate local operation, one can proceed to find $X_n$, $Y_n$ as given in Eq.~\eqref{eqn:rec}.
% \end{comment}

\end{document}